# SISTEM PENDUKUNG KEPUTUSAN PEMBERIAN BEASISWA BIDIK MISI

**Pesos Umami[1], Leon Andretti Abdillah[2*], Ilman Zuhri Yadi[3]**

[1,2,3] Program Studi Sistem Informasi, Fakultas Ilmu Komputer, Universitas Bina Darma
[3] Jl. Ahmad Yani No. 12, Plaju, Palembang, 30264
[1] pesosumami@ymail.com, *[2] leon.abdillah@yahoo.com, [3] ilmanzuhriyadi@mail.binadarma.ac.id

## Abstract

Decision support systems (DSS) have been used in many applications to speed up decision making. This article will discuss the implementation of DSS for decision-making for scholarships "bidik misi". In terms of processing data related to the bidik misi scholarship are accordance with the regulations set by UBD. To facilitate the selection process of bidik misi scholarship then the authors used SAW method. The criteria for this scholarship involves academic achievement index, parents income, parents dependents number, semester, etc. Author also used waterfall model approacah to developed the systems. The result is a decision support system that able to help decisions making quickly and in accordance with the rules of bidik misi scholarship.

**Kata kunci** : *Sistem Penunjang Keputusan (SPK), Simple Additive Weighting (SAW), Bidik Misi.*

## 1. Pendahuluan

Sistem penunjang keputusan (SPK) atau *Decision support systems* (DSS) merupakan salah satu bagian dari sistem informasi yang telah banyak diterapkan untuk memudahkan pengambilan keputusan baik untuk jangka pendek, menengah, ataupun panjang. Sejumlah keputusan yang diambil tidak saja berhubungan dengan aktivitas bisnis semata, namun juga dapat berhubungan dengan bidang-bidang lain, seperti bidang pendidikan misalnya.

Pendidikan merupakan suatu kebutuhan primer yang sejak dini hingga dewasa hendaknya dirasakan oleh seluruh masyarakat. Hal ini sesuai dengan amanat UUD Negara Kita, anjuran agama, dan menjadi penentu kemajuan suatu bangsa [1]. Pendidikan juga merupakan variabel vital untuk pembangunan suatu bangsa. Suatu bangsa bisa maju dengan cepat dibandingkan dengan negara lain karena penyebaran pengetahuan (*knowledge*) yang merata keseluruh lapisan masyarakatnya. Institusi yang paling bertanggung jawab untuk penyebaran pengetahuan adalah institusi pendidikan [2]. Dalam upaya pembangunan bidang pendidikan, pemerintah selaku institusi utama pendorongnya telah melakukan sejumlah kebijakan, seperti 1) wajib belajar 9 tahun, pengadaan beasiswa-beasiswa, program bidik misi [3] di perguruan tinggi, dll.

Pemberian beasiswa merupakan program kerja yang ada di setiap Universitas atau Perguruan Tinggi. Program beasiswa diadakan untuk meringankan beban mahasiswa dalam menempuh masa studi, khususnya dalam masalah biaya. Pemberian beasiswa dilakukan secara selektif sesuai dengan jenis beasiswa yang diadakan. Banyak sekali beasiswa yang ditawarkan kepada mahasiswa yang berprestasi dan yang kurang mampu. Salah satu beasiswa yang ditawarkan Universitas Bina darma yaitu beasiswa bidik misi. Lembaga tersebut mengeluarkan beasiswa setiap tahun bagi mahasiswa.

Bidik Misi memberikan beasiswa kepada mahasiswa Universitas Bina Darma Palembang. Sesuai dengan peraturan yang telah ditentukan oleh Bidik Misi pada UBD Palembang untuk beasiswa, maka diperlukan kriteria-kriteria untuk menentukan siapa yang akan dipilih untuk menerima beasiswa. Kriteria dalam studi ini adalah sesuai indeks prestasi akademik, penghasilan orangtua, jumlah tanggungan orangtua, semester dan lain-lain. Oleh sebab itu tidak semua calon pengajuan beasiswa tersebut diterima, hanya yang memenuhi kriteria saja yang akan menerima beasiswa tersebut. Pengajuan beasiswa Bidik Misi pada UBD cukup banyak serta indikator dalam penyeleksian berkas pengajuan beasiswa yang masih secara manual. Dengan sistem yang ada sekarang Bidik Misi pada UBD sangat sulit untuk menentukan siapa yang layak menerima beasiswa tersebut, karena banyaknya pengajuan beasiswa dan banyaknya kriteria-kriteria yang harus ditentukan untuk menentukan siapa yang benar-benar berhak mendapatkan beasiswa tersebut. Dengan demikian dibutuhkan sistem yang dapat membantu membuat keputusan penerima beasiswa dengan cepat dan tepat, untuk meringankan kerja bagian kemahasiswaan dalam menentukan penerima beasiswa.

Sejumlah penelitian telah dilakukan yang berhubungan dengan *decison making*, dan Bidik Misi. Penelitian tersebut, antara lain: 1) Wibowo [4] membangun SPK penentu penerima beasiswa BRI di FTI UII, 2) Antoni [5] menggunakan peubah penjelas: kepemilikan prestasi, jenis kelamin, lokasi asal SLA, status asal SLA, jumlah tanggungan orang tua, penghasilan orang tua, akreditasi SLA dan usia saat masuk IPB, 3) Afyanti [6] membahas Pemberian Kelayakan Kredit Pinjaman pada BRI Unit Segiri Samarinda, 4) Baruadi [7], membahas keterampilan sosial para penerima beasiswa bidik misi, dan 5) Arifin [8] menyurvei penggunaan beasiswa yang dikelola oleh mahasiswa penerima beasiswa bidik Misi.

Pada penelitian ini, panulis memanfaatkan *Fuzzy Multiple Attribute Decision Making* (FMADM). FMADM untuk digunakan dalam penelitian ini dimana langkah penyeleksian alternatifnya lebih pendek namun akan tetap menghasilkan keputusan optimal dalam menentukan alternatif terbaik dari berbagai alternatif berdasarkan kriteria tertentu [9]. Sedangkan *Simple Additive Weighting* (SAW) yaitu mencari penjumlahan terbobot dari rating kinerja pada setiap alternatif pada semua atribut. Metode SAW ini dipilih karena lebih efektif, lebih mudah pada proses perhitungan dalam penyeleksian penerima beasiswa dan lebih efisien [10]. Metode perangkingan diatas diharapkan akan memberikan penilaian yang lebih tepat. Hal ini dikarenakan penilaian didasarkan pada nilai kriteria dan bobot yang sudah ditentukan terlebih dahulu. Sebagai konsekuensinya penentuan penerima beasiswa lebih akurat.

Bagian selanjutnya dari artikel ini akan mengulas metode penelitian, hasil pembahasan, serta simpulan.

## 2. Metode Penelitian

### 2.1 Metode Pengembangan Sistem

Metode pengembangan sistem yang digunakan adalah *waterfall model* atau *classic life cycle* [11], yang terdiri dari: 1) *Communication: project initiation requirements gathering*, 2) *Planning: estimating scheduling tracking*, 3) *Modeling: analysis design*, 4) *Construction: code test*, dan 5) *Deployment: delivery support feedback* (Gambar 1).

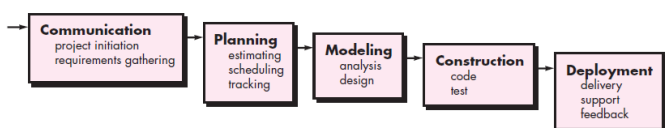

Gambar 1. *Waterfall model* atau *classic life cycle*

### 2.2 Analisis Penerimaan Beasiswa Bidik Misi UBD

Dalam penelitian ini, proses penyeleksian calon penerima bidik misi UBD mengikuti urutan logika berikut:

1) Bobot. Penerapan metode FMADM – SAW dalam penelitian ini memerlukan bobot dan kriteria untuk menentukan siapa yang akan terseleksi sebagai penerima beasiswa. Adapun kriterianya adalah: a) $C_1$ = Nilai, b) $C_2$ = Penghasilan Orangtua, c) $C_3$ = Jumlah Tanggungan Orangtua, dan d) $C_4$ = Semester. Dari masing-masing bobot tersebut, maka dibuat suatu variabel-variabelnya. Dimana dari suatu variabel tersebut akan dirubah kedalam bilangan *fuzzy*. Dibawah ini adalah *fuzzy* dari bobot: a) Sangat Rendah (SR) = 2, b) Rendah (R) = 4, c) Cukup Tinggi (CT) = 6, d) Tinggi (T) = 8, dan e) Sangat Tinggi (ST) = 10.

2) Kriteria Nilai. Kriteria nilai merupakan persyaratan yang ditentukan untuk pengambilan keputusan, berdasarkan jumlah nilai yang diperoleh oleh mahasiswa selama studi berlangsung. Interval nilai yang telah dikonversikan dengan bilangan *fuzzy* sbb: a) Nilai < 40 = 2, b) Nilai 40-60=4, c) Nilai 60-69=6, d) Nilai 70-84=8, dan e) Nilai 85-100=10.

3) Kriteria Penghasilan Orangtua. Kriteria penghasilan orangtua merupakan persyaratan yang ditentukan untuk pengambilan keputusan, berdasarkan jumlah penghasilan bulanan baik itu penghasilan tetap maupun tidak tetap. Interval nilai yang telah dikonversikan dengan bilangan *fuzzy* sbb: a) Penghasilan < 1 jt = 10, b) Penghasian 1 jt s.d < 2,5 jt = 8, c) Penghasilan 2,5 jt s.d. < 5 jt = 6, dan d) Penghasilan > 5 jt = 4.

4) Kriteria Jumlah Tanggungan Orangtua. Kriteria jumlah tanggungan orangtua merupakan persyaratan yang ditentukan untuk pengambilan keputusan, berdasarkan jumlah anak yang masih menjadi tanggungan orangtua berupa biaya hidup. Interval kelas yang telah dikonversikan dengan bilangan *fuzzy* sbb: a) Tanggungan 1 anak = 2, b) Tanggungan 2 anak = 4, c) Tanggungan 3 anak = 6, d) Tanggungan 4 anak = 8, dan e) Tanggungan 5 anak = 10.

5) Kriteria Semester. Kriteria semester merupakan persyaratan yang ditentukan untuk pengambilan keputusan, berdasarkan semester pemohon. Interval semester yang telah dikonversikan dengan bilangan *fuzzy* sbb: a) Semester 2 = 2, b), Semester 3 = 4, c) Semester 4 = 6, d) Semester 5 = 8, dan e) Semester 6 = 10.

6) Vektor Bobot (W). Berdasarkan hasil wawancara pada bagian beasiswa pada Universitas Bina Darma Palembang memberikan bobot setiap kriteria sebagai berikut: a) $C_1$ (Nilai) = 40%, b) $C_2$ (Penghasilan orangtua) = 30%, c) $C_3$ (Tanggungan orangtua) = 10%, dan d) $C_4$ (Semester) = 20%.

### 2.3 Use Case Diagram

*Usecase* diagram memperlihatkan hubungan-hubungan yang terjadi antara aktor-aktor dengan *usecase* dalam sistem.

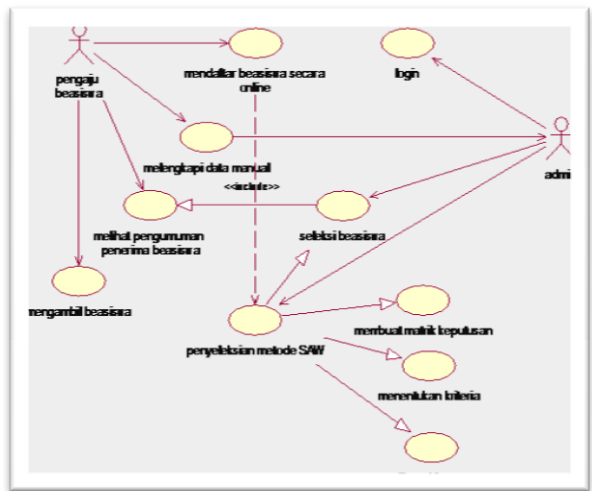

Gambar 2. *Use Case Diagram*

Berdasarkan gambar *usecase* diagram di atas menjelaskan bahwa pengaju beasiswa mendaftarkan beasiswa secara *online* setelah mendaftar pengaju melengkapi data dan menyerahkannya kebagian pengurus beasiswa di Universitas Bina Darma. Bagian pengurus beasiswa di Universitas Bina Darma Palembang melakukan peyeleksian mahasiswa dengan menggunakan metode SAW.

### 2.4 Rancangan *Database*

*Database* adalah kumpulan tabel yang saling terkait untuk menampung sejumlah data terkait agar program yang dibuat sesuai dengan apa yang diinginkan. Perancangan *database* [12] bertujuan: 1) untuk memenuhi informasi yang berisikan kebutuhan-kebutuhan user secara khusus dan aplikasi-aplikasinya, 2) memudahkan pengertian struktur informasi, dan 3) mendukung kebutuhan-kebutuhan pemrosesan dan beberapa obyek penampilan (*response time, processing time,* dan *storage space*).

*File-file* yang diperlukan dan akan digunakan di dalam sistem pendukung keputusan menggunakan FMADM (*Fuzzy Attribute Multiple Decision Making*) metode *Simple Additive Weighting* (SAW). Sejumlah tabel utama yang ada pada rancangan basisdata dapat dilihat pada tabel 1.

Tabel 1. Daftar Tabel Utama

| No | Tabel | Keterangan |
|---|---|---|
| 1 | Pemohon | menampilkan dan menyimpan data-data pemohon beasiswa |
| 2 | Penerima Beasiswa | menampilkan siapa penerima beasiswa berdasarkan kriteria yang telah ditentukan |
| 3 | Seleksi | menampilkan dan menyimpan data-data hasil dari proses seleksi penerimaan beasiswa |

## 3. Hasil dan Pembahasan

Berdasarkan hasil penelitian yang telah dilakukan pada Universitas Bina Darma Beasiswa Bidik Misi, maka didapatkan hasil akhir sebuah sistem yaitu Sistem pendukung Keputusan untuk menentukan penerima beasiswa menggunakan FMADM dengan metode SAW. Hasil ini didapt setelah menerapkan analisis dan perancangan ke dalam bahasa pemrograman *PHP* dan *MySQL*. Adapun aplikasi ini terdiri dari beberapa menu antara lain: 1) Menu Pendaftaran Beasiswa. Merupakan menu dimana untuk siswa menginputkan datanya yang nanti akan masuk ke dalam daftar pemohon beasiswa, 2) Menu Login. Merupakan menu dimana admin harus masuk terlebih dahulu sebelum melakukan penyeleksian, 3) Menu Periode Beasiswa. Merupakan menu dimana admin dapat melihat penerima beasiswa berdasarkan tahun yang diambil ataupun dapat melihat penerima beasiswa dari tahun sebelumnya, 4) Menu Daftar Pemohon. Merupakan menu dimana mahasiswa yang sudah mendaftar maka data-datanya akan langsung masuk ke dalam daftar pemohon, 5) Menu Seleksi Beasiswa. Merupakan menu dimana admin melakukan penyeleksian terhadap mahasiswa yang berhasil mendaftar beasiswa, 6) Menu Hasil Seleksi. Merupakan menu dimana admin melihat hasil penyeleksian dari proses seleksi beasiswa, dan 7) Menu Penerima Beasiswa. Merupakan menu dimana dapat melihat data-data yang menerima beasiswa.

### 3.1 Halaman Menu *Index*

Merupakan halaman utama ketika *user* membuka halaman utama, yang mempunyai komponen halaman diantaranya *header,* login, profil, visi misi, pendaftaran dan tentang beasiswa. Seperti yang terlihat pada gambar dibawah ini:

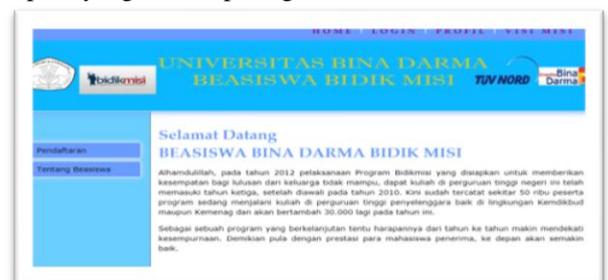

Gambar 3. Halaman Utama

### 3.2 Halaman Pendaftaran Pemohon

Halaman menu pendaftaran pemohon ini merupakan halaman yang akan diisi pemohon dalam mendaftar beasiswa secara *online* yang nantinya data-data tersebut akan masuk kedalam daftar calon penerima beasiswa, seperti gambar berikut ini.

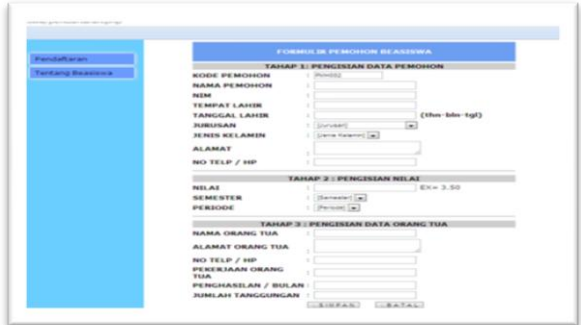

Gambar 4. Halaman Pendaftaran Pemohon

### 3.3 Menu Periode Beasiswa

Halaman menu periode beasiswa ini merupakan halaman yang dapat melihat data-data penerimaan beasiswa setiap tahunnya, seperti gambar dibawah ini.

Gambar 5. Halaman Periode Beasiswa

### 3.4 Menu Daftar Pemohon

Halaman menu daftar pemohon ini merupakan halaman untuk menampung nama-nama mahasiswa yang mendaftar beasiswa secara keseluruhan, seperti gambar 62.

Gambar 6. Halaman Daftar Pemohon

### 3.5 Menu Seleksi Beasiswa

Halaman menu seleksi beasiswa ini merupakan halaman untuk melakukan penyeleksian terhadap mahasiswa yang sudah mendaftar beasiswa dan memenuhi kriteria yang telah ditentukan, pada menu ini dilakukan seleksi dengan proses FMADM kemudian proses SAW, seperti gambar dibawah ini.

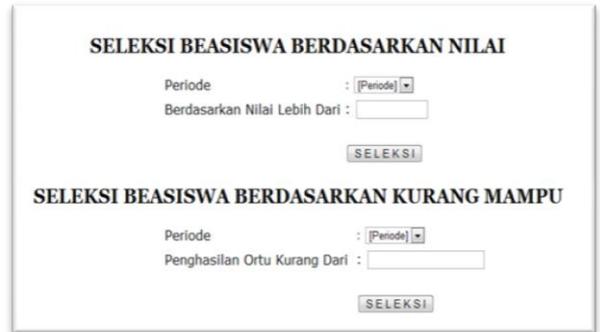

Gambar 7. Halaman Seleksi Beasiswa

### 3.6 Proses Seleksi FMADM

Halaman menu proses seleksi FMADM ini merupakan halaman untuk memproses data-data siswa berdasarkan kriteria yang telah ditentukan dengan metode *Fuzzy Multiple Attribute Decision Making* (FMADM) yang nanti akan dilanjutkan ke proses *Simple Additive Weighting* (SAW), seperti gambar dibawah ini.

Gambar 8. Halaman Proses Seleksi FMADM

### 3.7 Menu Hasil Seleksi

Halaman menu hasil akhir seleksi beasiswa dengan metode SAW ini merupakan halaman yang dilakukan setelah melakukan proses FMADM. Yang mana pada proses SAW ini langsung dikalikan dengan vector bobot yang sudah ditentukan dari masing-masing kriteria, seperti gambar dibawah ini.

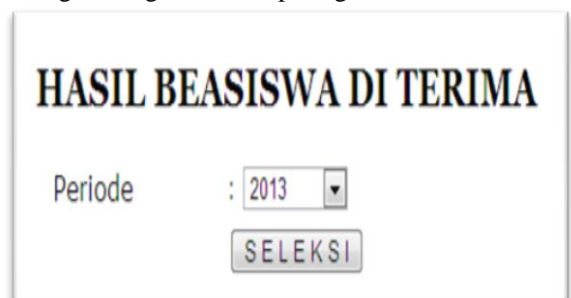

Gambar 9. Halaman Hasil Seleksi

### 3.8 Menu Penerima Beasiswa

Halaman menu penerima beasiswa ini merupakan halaman untuk data-data siswa yang menerima beasiswa berdasarkan jenis-jenis beasiswa dan periode beasiswa pada Universitas Bina Darma Beasiswa Bidik Misi Palembang, seperti gambar dibawah ini.

Setelah dimasukkan periode dan jenis beasiswa maka akan tampil hasil penerima beasiswa berdasarkan periode dan jenis beasiswa, seperti gambar dibawah ini.

Gambar 10. Halaman Penerima Beasiswa

4. **Simpulan**

Berdasarkan uraian pada bagian-bagian artikel di atas, penulis dapat mengambil sejumlah simpulan, sebagai berikut:

a. Sistem yang dibangun untuk mengolah data pemohon beasiswa menjadi informasi yang dapat digunakan oleh Universitas Bina Darma dalam pengambilan keputusan untuk menentukan penerima beasiswa.

a. Sistem Penunjang Keputusan yang dibangun ini dapat mempercepat proses penyeleksian beasiswa, karena proses seleksi dilakukan secara otomatis. Dimana mahasiswa mengisi data-data pada formulir pendaftaran sicara *online* dengan benar dan sesuai dengan ketentuan yang ada selanjutnya data mahasiswa tersebut akan langsung masuk dalam sistem untuk tahap penyeleksian dengan metode FMADM dan SAW sehingga mendapatkan hasil penerima beasiswa dari hasil proses sistem tersebut.